\documentclass[referee,useAMS,usenatbib,onecolumn]{mn2e}

\usepackage{graphicx}
\usepackage{url}
\usepackage{epsfig}
\usepackage{color,amsmath,amssymb}

\renewcommand{\arraystretch}{0.8}

\newcommand{\4}{$_{4}$}
\newcommand{\cm}{cm$^{-1}$}

\newcommand{\lnr}{^{\ell}}
\newcommand{\p}{^\prime}
\newcommand{\pp}{^{\prime\prime}}

\newcommand{\Td}{${\mathcal T}_{\rm d}$}

\newcommand{\ai}{\textit{ab initio}}

\newcommand{\schr}{Schr\"{o}dinger}

\title{ExoMol line lists IV: The rotation-vibration spectrum of methane
up to 1500~K}

\date{\today}
\author[Yurchenko and Tennyson]{\large
Sergei N. Yurchenko and Jonathan Tennyson \\
Department of Physics and Astronomy, University College London, London WC1E 6BT, UK}
\date{Accepted XXXX. Received XXXX; in original form XXXX}

\pagerange{\pageref{firstpage}--\pageref{lastpage}} \pubyear{2014}

\begin{document}

\maketitle

\begin{abstract}

A new hot line list is calculated for $^{12}$CH\4\ in its ground
electronic state. This line list, called 10to10, contains 9.8
billion transitions and should be complete for temperatures up to
1500~K. It covers the wavelengths longer than 1~$\mu$m and includes
all transitions to upper states with energies below $hc \cdot
18\,000$~\cm\ and rotational excitation up to $J=39$.  The line list
is computed using the eigenvalues and eigenfunctions of CH\4\
obtained by variational solution of the \schr\ equation for the
rotation-vibration motion of nuclei employing program TROVE.  An
{\it ab initio} dipole moment surface and a new `spectroscopic'
potential energy surface are used.  Detailed comparisons with other
available sources of methane transitions including HITRAN,
experimental compilations and other theoretical line lists show that
these sources lack transitions both higher temperatures and near
infrared wavelengths.  This line list is suitable for modelling
atmospheres of cool stars and exoplanets. It is available from the
CDS database as well as at \url{www.exomol.com}.

\end{abstract}

\label{firstpage}

\section{Introduction}

Methane plays an important role in atmospheric and astrophysical
chemistry. Its rotation-vibration spectrum is of key importance for
models of the atmospheres of bodies ranging from Titan to brown
dwarfs. Any temperature-dependent model of the methane spectrum
requires comprehensive information on the associated transition
intensities.

Methane was detected in the exoplanet HD189733b by \citet{08SwVaTi.exo} and
further studied using ground-based observations by \citet{10SwDeGr.exo}.
However the abundance of exoplanetary methane has sometimes proved
controversial \citep{10StHaNy.exo,jt495}. On earth methane is an important
global warming species \citep{04RoDoxx.CH4} and
a biosignature \citep{93SaThCa.CH4};  Spectra of hot methane are also required for a variety of terrestrial
applications including the study of combustion \citep{07JoGaCh.CH4} and sensing \citep{13WoRhBr.CH4}.
Methane  has been detected on Mars  \citep{07AtMaWo.CH4} although these observations have
not always proved repeatable \citep{12Krxxxx.CH4}. On earth-like planets
  outside our solar system, whose spectra of course have yet to be
  observed, methane has long been thought of as a key potential
  biosignature \citep{93SaThCa.CH4,07AtMaWo.CH4}.

Methane is well known in cool carbon stars and brown dwarfs
  \citep{09Bexxxx.exo}; indeed T-dwarfs are often referred to as
    methane dwarfs \citep{jt484}. However modelling methane in these objects with
    available laboratory data has often proved difficult
    \citep{96GeKuWi.CH4,12BaKe}. Similarly hot methane emissions were observed
    in aftermath of the collision of comet Shoemaker-Levy 9 with
    Jupiter \citep{95MaDrBe.CH4,jt198}.

Given the importance of methane spectra, a growing number of
$^{12}$CH$_4$ rotation-vibration line lists are available. The standard for
atmospheric spectroscopy is HITRAN and the recently released 2012
edition \citep{jt557} contains a thorough update for methane as
detailed by \citet{13BrSuBe.CH4}. However, HITRAN is designed for work
at Earth atmosphere temperatures and its companion, high-temperature
database HITEMP \citep{jt480} does not include methane. Other sources
include a high temperature line list computed  by \citet{09WaScSh.CH4},
the MeCaSDa spectroscopic database containing largely empirical line list
\citep{13BaWeSu.CH4} and a
partial experimental line list WKMC constructed by
\citet{13CaLeMo.CH4}. Finally high-temperature experimental line lists
are available due to
\citet{03NaBexx.CH4}, \citet{08ThGeCa.CH4}, and \citet{12HaBeMi.CH4}.
Measured methane cross sections are also provided in the PNNL
database~\citep{PNNL}. As discussed below, none of these compilations
are complete at the elevated temperatures considered in this work.

Recently we \citep{jt528} started the ExoMol project which aims to
generate a comprehensive library of line list for all molecules likely
to be important for modelling atmospheres of exoplanets and cool
stars.  Here we employ the variational program TROVE
\citep{07YuThJe.method} to construct a synthetic line list for
$^{12}$CH\4\ in its ground electronic state. To this end the \schr\
equation for the rotation-vibration motion of nuclei is solved to
obtain eigenvalues (ro-vibrational energies) and eigenfunctions (nuclear
motion wavefunctions). The
latter are necessary for ro-vibrational averaging of the dipole moment
of the molecule and thus to compute the transitional probabilities,
usually expressed in terms of the Einstein coefficients or line
strengths. Such calculations are based on the use of a potential
energy surface (PES) and dipole moment surfaces (DMSs).

There have been many quantum-chemical studies of methane. We will
concentrate on those aimed at producing comprehensive
rotation-vibration line lists.  \citet{06OyYaTa.CH4} presented \ai\
PES calculations based at the CCSD(T)/cc-pVTZ and MP2/cc-pVTZ level of
theory. \citet{09WaScSh.CH4} used RCCSD(T)/aug-cc-pVTZ theory to compute
an hot line list containing 1.4 millions.
\citet{11NiReTy.CH4} constructed a PES which they showed to provide
accurate (within 1 \cm) vibrational energies of CH\4.
\citet{13NiReTy.CH4} recently presented new \textit{ab
  initio},  methane DMSs based on the CCSD(T)/cc-pCVQZ level of
theory.  \citet{13ReNiTy.CH4,13ReNiTy.CH4.i} employed these PES and DMSs to simulate the room
temperature infrared (IR) absorption spectrum of CH\4. Comparison with laboratory experiments showed that their
predicted intensities agreed well indicating the high-quality of the
underlying \ai\ DMSs. Other recent studies
on methane include those by \citet{01ScPaxx.CH4} and \citet{02Scxxxx.CH4},
\citet{13WaCaxx.CH4} and \citet{13MiChTr.CH4}.



In this work we present a new `spectroscopic' PES obtained by
fitting an \ai\ PES from \citet{jt555} to the experimental energies of
CH\4\ with $J=0,1,2,3,4$. The \ai\ dipole moment surface (DMS)
of \citet{jt555}, obtained at the CCSD(T)/aug-cc-pVTZ level of theory,
is used to compute the line strengths.

The paper is structured as follows. The following section introduces
the PES and DMSs, as well as the quantum number scheme used to label
the ro-vibrational states of CH\4 (also discussed in detail in the
appendix).  Section \ref{s:variational} describes the variational
procedure used to solve the \schr\ equation. The intensity calculations and the
line list generated are detailed in Section~\ref{s:linelist}. The
evaluation of the new line list and comparisons against different
empirical and theoretical line lists are given in
Section~\ref{s:analysis}. Some conclusions are offered in
Section~\ref{s:conclusion}.

\section{Background to the calculation}
\label{s:qn}

In the following CH\4\ and methane will refer to the main isotopologue
$^{12}$CH\4\ of methane unless specified.

Methane is a symmetric five atomic molecule characterised by nine
vibrational degrees of freedom with a vanishing permanent dipole
moment.  It is a very high symmetry molecule of the \Td(M) symmetry
group containing a number of degenerate modes. As discussed in our
recent work on ammonia \citep{jt546}, this situation means that some
care has to be taken in selecting an appropriate set of quantum numbers.
Below we specify our chosen quantum numbers for CH\4;
reasons for this choice are given in the appendix.

The symmetry properties of methane spectra have been carefully considered
by \citet{01Hoxxxx.CH4} and we used this work as our starting point.
Using the Molecular Symmetry group
\citep{04BuJexx.method} the total ro-vibrational states of CH\4\ have
total symmetry, $\Gamma_{\rm tot}$, either $A_1$, $A_2$, $E$, $F_1$, or $F_2$.
Considering the H nuclear spin, the nuclear statistical weight,
$g_{\rm ns}$,  takes the
value 5, 5, 2, 3, and 3 for each symmetry respectively.
The electric dipole transitions obey the following symmetry-determined
selection rules:
\begin{equation}\label{e:selection-rules}
  A_1 \leftrightarrow A_2 \;, E \leftrightarrow E \; , F_1 \leftrightarrow F_2
\end{equation}
with the standard rotational angular momentum, $\mathbf J$, selection rules:
\begin{equation}\label{e:selection-rules:J}
  J \leftrightarrow J\pm 1 \;\; , J\p+J\pp \ne 0 .
\end{equation}
Our set of  quantum numbers consists of the following 15 labels:
\begin{equation}\label{e:qns}
  {\rm QN} = \Gamma_{\rm tot}, J, K, \Gamma_{\rm rot}, n_1, n_2, L_2, n_3,L_3, M_3, n_4,L_4,M_4, \Gamma_{\rm vib}, N_{J, \Gamma_{\rm tot}},
\end{equation}
where $\Gamma_{\rm rot}$ and $\Gamma_{\rm vib}$ are the symmetries of the
rotational and vibrational wavefunction respectively, $K = 0, \ldots J-1, J$ is the absolute value of the projection 
of the rotational angular momentum of $J$ onto the body-fixed $z$ axis, and $n_1, n_2, L_2, n_3, L_3, M_3, n_4,
L_4, M_4$ are the normal mode vibrational quantum numbers (see the mode designation in
Table~\ref{t:irreps}). We follow \citet{jt546} and use the only the
absolute values of the vibrational angular momentum quantum numbers
$L_i = n_i, n_i -2, \ldots, 0 (1)$. $M_i \le L_i$ is a multiplicity index used to count states within a given $n_i, L_i$ set (see \citet{06BoReLo.CH4}) and  indicate the symmetry according to Table~\ref{t:F:M}.
Finally as it is often not possible to uniquely define CH\4\
quantum numbers, a counting number $N_{J, \Gamma_{\rm tot}}$ is also included. This number
runs over states of a given total symmetry and $J$.



\begin{table}
\caption{}
\label{t:irreps} \footnotesize
\begin{center}
\begin{tabular}{cccl}
  \hline
  Mode & Symmetry && Type \\
  \hline
    1  & $A_1$  && symmetric stretch \\
    2  & $E  $  && asymmetric bend \\
    3  & $F_1$  && asymmetric stretch \\
    4  & $F_2$  && asymmetric bend \\
  \hline
\end{tabular}
\end{center}
\end{table}

\begin{table}
\caption{ The classification of the multiplicity index $M_i$ ($i=3,4$, $M_i\le L_i$) by symmetry
($n=1,2,3\ldots$) }
\label{t:F:M}  \footnotesize
\begin{center}
\begin{tabular}{cc}
  \hline
   Symmetry  & M  \\
  \hline
   $A_1$ & 0 + 12 n  \\
   $A_2$ & 6 + 12 n  \\
   $E  $ & 2 + 6 n  \\
   $F_1$ & 3 + 4 n  \\
   $F_2$ & 1 + 4 n  \\
  \hline
\end{tabular}
\end{center}
\end{table}


To calculate the ro-vibrational energies in the Born-Oppenheimer
approximations, a PES is required.  In this work we use a new
`spectroscopic' PES obtained by refining an \ai\ PES through fits to
experimental ro-vibrational energies. To this end a set of
experimental term values with $J\le 4$ was selected ranging up to
$6100$ \cm\ derived from frequencies collected in the HITRAN~2008
database \citep{jt453}. We recently produced a `spectroscopic' CH\4\ PES
using a similar procedure by fitting an \ai\ CCSD(T)-F12/aug-cc-pVQZ
PES to the same set of experimentally derived energies
\citep{jt555}.  Here the `spectroscopic' PES of \citet{jt555} was
further refined.  This was necessary because our refinement procedure,
and therefore the resulting PES, is sensitive to the size of the basis
set used in the variational calculations. The basis set employed here
differs from that used by \citet{jt555}, see below for more details. The resulting PES is
given as Supplementary Material to this paper in the form of a Fortran
95 program. It should be noted, however, that this PES has an
`effective' character and guarantees to give accurate results only in
conjunction with the same method and basis set used to produce it
with. This obvious disadvantage is a result of the non-converged basis
set used in TROVE, as well as the approximate character of the kinetic
and potential energy expansions employed.

The fitting set of our `experimentally-derived' term values (\cm) was
built from combination differences arising from the transition
wavenumbers collected in HITRAN~2008. It should be noted that only a
few of the term values were supported by more than one
transition.

In our procedure, as described in detail by \citet{11YuBaTe.method},
the refined PES is formulated as a correction to the original PES:
\begin{equation}\label{e:V}
  V = V_0 + \Delta V,
\end{equation}
where both $V_0$ and $\Delta V$ are presented as expansion in terms of
the internal coordinate displacements. At the first stage, the
Schr\"{o}dinger equations are solved for the initial potential $V_0$
to obtain the eigenfuncitons $\Psi_{J,\lambda}^{(0)}$ for a set of the
rotational quantum numbers $J$. These eigenfunctions are then used as
basis functions for solving the Schr\"{o}dinger equations with the
probe PES $V$ during the fitting procedure. In this way only the
matrix elements $\langle \Psi_{J,\lambda} | \Delta V |
\Psi_{J,\lambda\p} \rangle $ are required when solving the
corresponding Schr\"{o}dinger equations variationally.

\begin{figure}
\caption{Residuals of CH\4\ experimentally-determined and calculated energy term values.
Calculated using the refined PES before the shift of the vibrational band  centers  (red empty circles)  and after (blue filled circles), see text.  }
\centering
{\leavevmode \epsfxsize=11.0cm \epsfbox{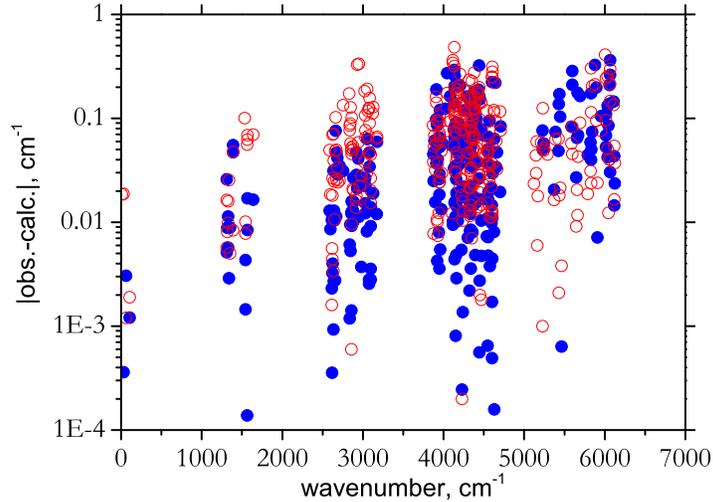}
\label{f:obs-calc:2}}
\end{figure}

The accuracy achieved in this fit is illustrated
in Fig.~\ref{f:obs-calc:2}, where the absolute values of the
obs.-calc. residuals (\cm) are shown. The
total root mean square (rms) error for 317 term values used is 0.11
\cm\ with the major data points distributed between 0.01 and 0.1 \cm.
It should be noted however that some of the transitions present in
HITRAN~2008 (especially above 5\,000~\cm) appeared to be relatively
large outliers and therefore could not be included into our fitting
set. This may indicate problems with our PES for the vibrational modes
not properly sampled by out fitting set and/or assignment problems
in HITRAN~2008.


\section{Variational calculations}
\label{s:variational}

The high symmetry of methane meant that a specially adapted version of
TROVE had to be developed, details of which are given by
\citet{jt555}.  In TROVE the basis set is built using the polyad
number, which for CH\4\ is given by
\begin{equation}\label{e:polyad}
  P = 2 (v_1 + v_2 + v_3 + v_4) + v_5 + v_6 + v_7 + v_8 + v_9 ,
\end{equation}
where $v_i$ is a vibrational quantum number associated with the
one-dimensional primitive basis function $\phi_{v_i}(\xi_{i}\lnr)$
($i=1,2,3,\ldots,9$) and $\xi_{i}\lnr$ is an internal linearized coordinate, see below.
In the conventional normal mode representation this condition corresponds to
\begin{equation}\label{e:polyad:norm}
  P = 2 (n_1 + n_3) + n_2 + n_4.
\end{equation}
The stretching basis functions $\phi_{v_i}(\xi_{i}\lnr)$ ($i=1,2,3,4$)
were obtained by numerically solving a reduced, one-dimensional
Hamiltonian problem using the Numerov-Cooley method. Harmonic
oscillators are used for the bending basis functions $i=5,6,7,8,9$.
The basis set used was constructed from two major contributions, (i)
all basis functions with the primitive quantum numbers satisfying $P
\le 10$ and (ii) stretching functions ranging up to $P=20$ but with
some high $P$-polyad ($P\ge 17$) stretching contributions that couple
all modes together removed.  This scheme is designed to reduce the
basis set to the manageable size and at the same time to retain
stretching basis functions with higher excitations than bending,
giving preference to less-mixed modes. Our underlying assumption for this
is that the stretching excitations carry the strongest transitions.

Following \citet{jt555}, the final contracted basis  functions were obtained employing the $J=0$ representation.  In
this representation the $J=0$ (vibrational) eigenfunctions are used
together with symmetrised rigid rotor functions to construct the basis set
functions for $J>0$. For each $J$ value, five symmetrised
Hamiltonian matrices employing the $J=0$ contraction scheme
are computed  and then diagonalized. In order to improve the prediction property of the refined potential of CH\4\ (see Fig.~\ref{f:obs-calc:2}) further,
artificial band centre shifts were added to the $J=0$ energies following the empirical basis set correction scheme (EBSC) \citep{jt466}. The result of the improvement can be see in Fig.~\ref{f:obs-calc:2} as well.

In the TROVE approach the Hamiltonian operator is represented as an
expansion around a reference geometry taken here at the molecular
equilibrium. The kinetic energy operator is expanded in terms of the
nine coordinates $\xi_{i}\lnr$, which are linearized
versions of the internal coordinates $\xi_i$ ($i=1,\ldots,9$), see
\citep{jt555}; the potential energy function is expanded in terms of
$1-\exp(-a \xi_i\lnr)$ ($i=1,2,3,4$) for the four stretching and $\xi_j\lnr$
($j=5,6,7,8,9$) for the five bending modes. The
latter expansions were applied to the refined PES introduced above.
The kinetic and potential energy parts were truncated at sixth-
and eighth-order, respectively.

\begin{figure}
\caption{Size of the $F_1$-symmetry Hamiltonian matrix and the number of eignevalues below 18000~\cm\ for each $J$. }
\centering
{\leavevmode \epsfxsize=11.0cm \epsfbox{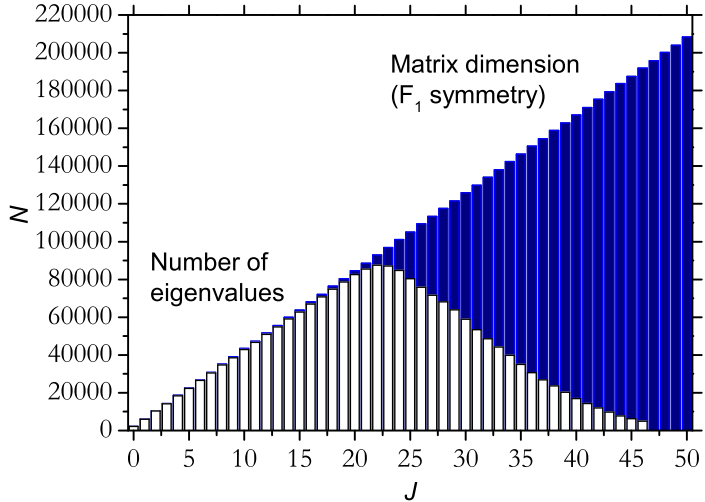}
\label{f:nn:J}}
\end{figure}

The dimensions of the $J=0$ basis sets used are 837, 585, 1\,418, 1\,916 and
2\,163 for the $A_1$, $A_2$, $E$, $F_1$, and $F_2$ symmetries,
respectively. The (symmetrically adapted) ro-vibrational basis
functions are represented by the (symmetrically reduced) direct
product of $(2J+1)$ rigid rotor wavefunctions and the $J=0$
eigenfunctions. The size of the ro-vibrational basis set scales
linearity with $J$ as illustrated in Fig.~\ref{f:nn:J}, where
the size of the $F_{1x}$ matrix is shown for $J=0 \ldots 39$.  The
largest matrix to be diagonalized ($F_{1x}$, $J=39$) has $163\,034 \times
163\,034$ elements. The dimension of the matrices of different symmetries
scale approximately in the ratio 1:1:2:3:3 for the $A_1$, $A_2$, $E$, $F_{1}$,
$F_{2}$ symmetries, respectively.

\begin{figure}
\caption{Number of energy levels and lines in the 10to10 line list per the rotational quantum number, $J$ .  }
\centering
{\leavevmode \epsfxsize=11.0cm \epsfbox{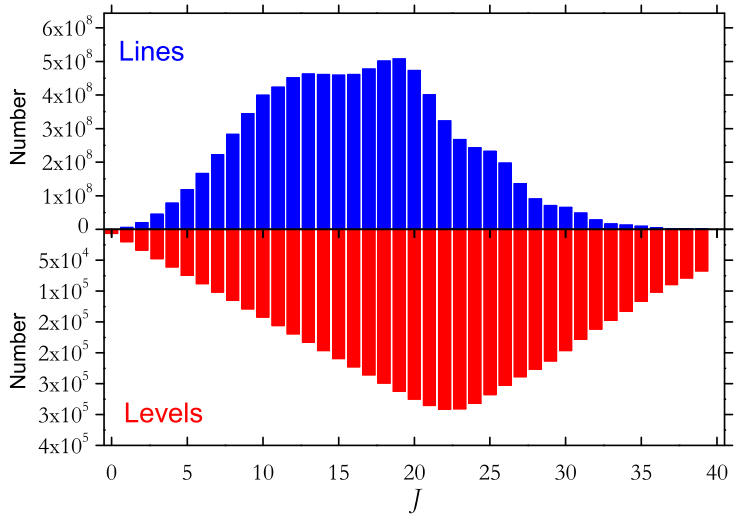}
\label{f:lines:J}}
\end{figure}

LAPACK routine DSYEV and ScaLapack routine PDSYEVD, as implemented in
the Intel MKL library, were used to diagonalize the Hamitlonian
matrices. The MPI version of LAPACK eigensolver PDSYEVD was used for
large matrices with $J \ge 24$. All eigen-roots were found
in the direct diagonalizations but only eigenvectors below the energy
threshold of $hc \cdot 18\,000$~\cm\ were stored and used in the line list
production to reduce the calculations and storage.
Fig.~\ref{f:lines:J} gives the total number of energy levels
and the number of transitions generated from these energy levels
as a function of $J$, summed over all five symmetries.
The linear dependence on $J$ at lower values is capped at
$J =21-22$ by the energy threshold.  At about
$J=50$ all energy levels of CH\4\  are above
18\,000~\cm. The total number of energy
levels subject to the thresholds $E_{\rm max} = 18\,000$~\cm\ and
$J_{\rm max}=39$ is 6~603~166.

\section{Line list calculations}
\label{s:linelist}

A spectroscopic line list is a catalogue of transitions containing
transition frequencies (or wavenumbers $\tilde\nu_{\rm if}$) and
transition line strengths (or Einstein A coefficients $A_{\rm if}$)
between a well-chosen set of  initial, $i$, and final, $f$, states.
A line list also necessarily  includes the total
state degeneracy as well as the lower state energy to allow
simulation of absorption or emission spectra at different temperatures.
It is also common to include the quantum numbers specifying states
involved in each transitions. For more details see \citep{jt511,jt548}.



The goal of this work is a comprehensive line list for CH\4\ covering
the wavenumber range up to 12\,000~\cm\ and applicable for
temperatures up to $T=1\,500$~K. To be complete
at $T=1\,500$~K, we estimated that the population is negligible for
energies  above 8\,000~\cm. This
defines our maximal value for lower state energy. We  find that
at $J=40$ all energy term values are above 8\,000~\cm. The lower
state energy threshold of $hc \cdot 8\,000$~\cm\ combined with the wavenumber
range $10\,000$~\cm\ defines the energy threshold for the upper state to
be $hc \cdot 18\,000$ ~\cm, which is also the maximal energy of CH\4\ included
in the final line list. That is, the thresholds for the energy and
eigenvectors are $J_{\rm max} = 39$, $E_{\max}\pp = hc\cdot  8\,000$~\cm,
$E_{\max}\p = hc\cdot 18\,000$~\cm. For the transition calculations we choose
an extended wavenumber range up to $\tilde\nu_{\rm max} = 12\,000$~\cm,
however the maximal temperature is applicable for the transitions up
to 10\,000~\cm\ (wavelengths longwards of 1~$\mu$m) only.



\begin{figure}
\caption{Number of intense lines as a function of intensity for different temperatures.
The $x$-axis gives the log of the intensity in cm/molecule,
while the $y$-axis represents the  number of transitions per each 10$^{x}$ cm/molecule bin.}
\centering
{\leavevmode \epsfxsize=11.0cm \epsfbox{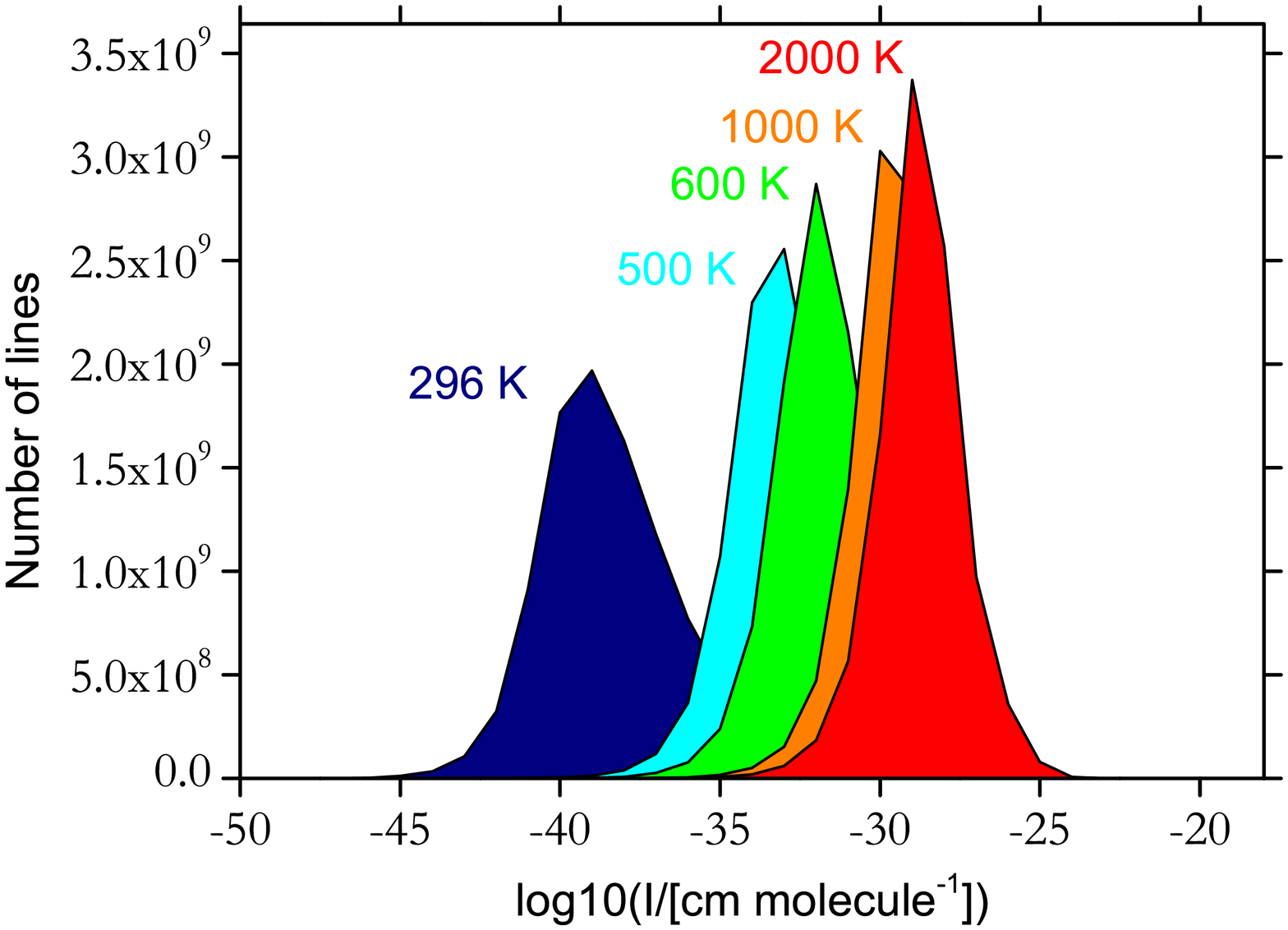}
\label{f:nlines:temp}}
\end{figure}

For the total number of levels considered, 6~603~166, the total number
of transitions computed is 9~819~605~160, i.e. almost 10$^{10}$, which is
why we call the line list `10to10'. Obviously not all these
transitions are important at all temperature. However, to
demonstrate that a large number of transitions do matter at elevated
temperatures, Fig.~\ref{f:nlines:temp} shows the density of lines
per an absorption intensity $I_{\rm if} = A \times 10^{x}$ unit
for different temperatures covering the whole wavenumber range $0\ldots
12\,000$~\cm.
The slightly deformed Gaussian-like distributions peak at $x = -39$,
$-33$, $-32$, $-30$, and $-29$ for $T=297$, 600, 1000, and 1500~K,
respectively, with long, very small tails spreading up to $x=-18$ which
are not visible at this scale. 98~\%\  of the $T=296$~K lines  we compute
have intensities smaller than $10^{-32}$ cm/molecule and barely
contribute to the total opacity at room temperature, and thus can be safely
ignored.  Conversely at $T=1500$~K, where the
absorption lines peak at $I_{\rm if} ~ 10^{-29}$ ~cm$/$molecule,
most of the hot lines (98~\%) appear to be stronger than $10^{-32}$~cm$/$molecule.
Many of these lines also overlap each other, due to the very high density of states,
which is not reflected on this figure.  In order to bring the line list to a more manageable size, we
could apply an intensity cutoff and compiled a reduced version of the
line list which gives absorption intensities higher than a certain threshold, say
$10^{-27}$~cm$/$molecule. It is known however that even weaker lines
may play important role for non-LTE environments \citep{jt330},
which is the reason for keeping even extremely weak
absorption transitions in 10to10. In fact there are
other, better ways of making this manageable such as cross sections
\citep{jt542} or $k$-coefficients \citep{96IrCaTa.method}.

\begin{figure}
\caption{Temperature dependence of the partition function of CH\4\   }
\centering
{\leavevmode \epsfxsize=11.0cm \epsfbox{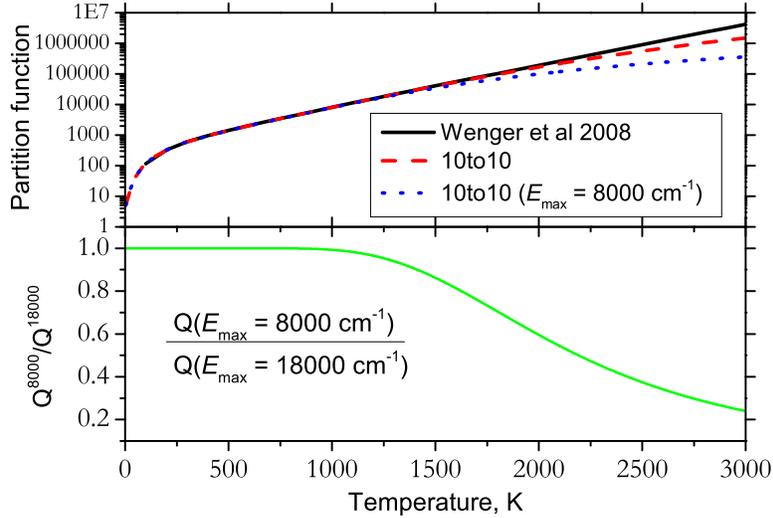}
\label{f:tempq}}
\end{figure}

\citet{08WeChBo.CH4} calculated the partition function for methane for
temperatures up to 3000~K by using all states up to the dissociation,
whose energies where modelled using different levels of
approximations.  Figure~\ref{f:tempq} (upper display) compares this
partition function with our values obtained by (a) summing all
energies from the 10to10 line list ($Q^{(18\,000)}$) and (b) summing
all energies below our lower cut-off of 8000~\cm\ ($Q^{(8\,000)}$).
This latter value is useful as it allows to estimate the completeness of the
10to10 line list as a function of temperature.  To
quantify this effect the lower display of Fig.~\ref{f:tempq} also
shows the ratio of the two partition functions, $Q^{(8\,000)}$ and
$Q^{(18\,000)}$.  As discussed above, our set of energies are the
subject of the threshold of $hc \cdot 18\,000$~\cm, which effects the
accuracy of our partition function at higher temperature.  At 1\,000~K,
10to10 appears to be complete but by 1\,500~K the 8\,000~\cm\
threshold leads to up to 15~\%\ under-sampling which increases rapidly
with temperature. Similar behaviour can be expected for the absorption
intensities when modelled at 1\,500~K and higher temperatures.

\section{The format of the line list}

We adopt the ExoMol format
\citep{jt548}
for the 10to10 line list.
In this format the line list consists of two files: (i) an Energy
file containing all information necessary to describe a given energy
level including, most importantly, the lower state energies and the
total degeneracies; (ii) a Transition
file, where the Einstein coefficients for all transitions are
specified. In the Energy file each energy level is numbered from 1 to
6~603~166, sorted by $J$ (from 0 to 39), symmetry ($A_1$, $A_2$, $E$,
$F_1$, and $F_2$), and energy. The counting number $i$, which is the
same as the row number, is an unique tag characterizing a given level.
The counting number is then followed by the lower state term value
(\cm) and the total statistical weights $g_i$, which for CH\4\ in its ground
($^1 \Sigma$) electronic state disregarding the hyperfine splitting, is
given by
\begin{equation}\label{e:tot-degen}
  g_i = (2J_i+1) g_{\rm ns}^{(i)}.
\end{equation}
Each energy record also includes the `good quantum
numbers' $J$ and $\Gamma_{\rm tot}$ (the total symmetry), as well as other quantum
numbers from Eq.~\eqref{e:qns}. An extract from the Energy file is
given in Table~\ref{t:Energy-file}.

The Transition file lists the electric dipole transition probabilities
in the form the Einstein $A$ coefficient $A_{\rm if}$ in s$^{-1}$ for each
transition i $\to$  f complemented by the $i$ and $f$ tags
identifying the states in the Energy file, for the lower (initial) and upper (final)
states, respectively.  For the convenience the Transition file is
split into 100~\cm\ wavenumber pieces (121 files). An extract from one
of the Transition files is given in Table~\ref{t:Transit-file}.
According to the ExoMol naming convention, the Energy file is called
10to10.states, while the Transition files are called
10to10-$nnnn$.transitions, where $nnnn$ indicates the
wavenumber region presented.

The ExoMol website also provides alternative tools allowing the user to
convert the line list into different representations, which currently
include HITRAN format \citep{jt557} and the cross-sections, see
\citet{jt542}. We also provide a spectrum
program  to simulate absorption and emission spectra using the 10to10
line list, in the form of the `stick' spectrum as well as of
cross-sections (with different broadenings).  The full line list can
be also downloaded from the Strasbourg Data Centre (SDC) , via
ftp://cdsarc.u-strasbg.fr/pub/cats/J/MNRAS/, or
http://cdsarc.u-strasbg.fr/viz-bin/qcat?J/MNRAS/.

\begin{table}
\caption{\label{t:Energy-file} Extract from the 10to10 Energy file.}
\scriptsize \tabcolsep=3pt
\renewcommand{\arraystretch}{1.0}
\begin{tabular}{rrrrrrrrrrrrrrrrrrrrrrrrrrrrrr}
    \hline
    \hline
  1     &        2        &  3  &  4  &  5  &  6  &  7  &  8  &  9  & 10  & 11  & 12  & 13  & 14  & 15  & 16  & 17  & 18  & 19  &   20   & 21  & 22  & 23  & 24  & 25  & 26  & 27  & 28  & 29  & 30  \\
    \hline
  $N$    &  $\tilde{E}$   & $g_{\rm tot}$&$J$&$\Gamma_{\rm tot}$&$n_1$&$n_2$&$L_2$&$n_3$&$L_3$&$M_3$ &$n_4$&$L_4$&$M_4$ &$\Gamma_{\rm vib}$&$J$&$K$&$\tau_{\rm rot}$&$\Gamma_{\rm rot}$& $N_{ J,\Gamma}$ &$|C_i|^2$&$v_1$&$v_2$&$v_3$&$v_4$&$v_5$&$v_6$&$v_7$&$v_8$&$v_9$\\
    \hline
  8836  &     1311.457042 & 15  &  1  &  2  &  0  &  0  &  0  &  0  &  0  &  0  &  1  &  1  &  1  &  5  &  1  &  1  &  1  &  4  &  1  &  1.00  &  0  &  0  &  0  &  0  &  0  &  0  &  1  &  0  &  0  \\
  8837  &     2632.988747 & 15  &  1  &  2  &  0  &  0  &  0  &  0  &  0  &  0  &  2  &  2  &  1  &  5  &  1  &  1  &  1  &  4  &  2  &  1.00  &  0  &  0  &  0  &  0  &  0  &  0  &  0  &  1  &  1  \\
  8838  &     2847.722094 & 15  &  1  &  2  &  0  &  1  &  1  &  0  &  0  &  0  &  1  &  1  &  1  &  5  &  1  &  1  &  1  &  4  &  3  &  1.00  &  0  &  0  &  0  &  0  &  1  &  0  &  1  &  0  &  0  \\
  8839  &     3028.725668 & 15  &  1  &  2  &  0  &  0  &  0  &  1  &  1  &  1  &  0  &  0  &  0  &  5  &  1  &  1  &  1  &  4  &  4  &  1.00  &  0  &  1  &  0  &  0  &  0  &  0  &  0  &  0  &  0  \\
  8840  &     3871.607969 & 15  &  1  &  2  &  0  &  0  &  0  &  0  &  0  &  0  &  3  &  1  &  1  &  5  &  1  &  1  &  1  &  4  &  5  &  1.00  &  0  &  0  &  0  &  0  &  0  &  0  &  3  &  0  &  0  \\
  8841  &     3955.888843 & 15  &  1  &  2  &  0  &  0  &  0  &  0  &  0  &  0  &  3  &  3  &  1  &  5  &  1  &  1  &  1  &  4  &  6  &  1.00  &  0  &  0  &  0  &  0  &  0  &  0  &  1  &  2  &  0  \\
  8842  &     4148.535610 & 15  &  1  &  2  &  0  &  1  &  1  &  0  &  0  &  0  &  2  &  2  &  1  &  5  &  1  &  1  &  1  &  4  &  7  &  1.00  &  0  &  0  &  0  &  0  &  1  &  0  &  0  &  1  &  1  \\
  8843  &     4223.887822 & 15  &  1  &  2  &  1  &  0  &  0  &  0  &  0  &  0  &  1  &  1  &  1  &  5  &  1  &  1  &  1  &  4  &  8  &  1.00  &  0  &  0  &  1  &  0  &  0  &  0  &  1  &  0  &  0  \\
  8844  &     4333.813607 & 15  &  1  &  2  &  0  &  0  &  0  &  1  &  1  &  1  &  1  &  1  &  1  &  5  &  1  &  1  &  1  &  4  &  9  &  1.00  &  0  &  1  &  0  &  0  &  0  &  0  &  1  &  0  &  0  \\
  8845  &     4353.438372 & 15  &  1  &  2  &  0  &  2  &  0  &  0  &  0  &  0  &  1  &  1  &  1  &  5  &  1  &  1  &  1  &  4  & 10  &  0.97  &  0  &  0  &  0  &  0  &  2  &  0  &  1  &  0  &  0  \\
  8846  &     4389.406791 & 15  &  1  &  2  &  0  &  2  &  2  &  0  &  0  &  0  &  1  &  1  &  1  &  5  &  1  &  1  &  1  &  4  & 11  &  0.97  &  0  &  0  &  0  &  0  &  0  &  2  &  1  &  0  &  0  \\
  8847  &     4555.544940 & 15  &  1  &  2  &  0  &  1  &  1  &  1  &  1  &  1  &  0  &  0  &  0  &  5  &  1  &  1  &  1  &  4  & 12  &  1.00  &  0  &  1  &  0  &  0  &  1  &  0  &  0  &  0  &  0  \\
  8848  &     5158.823917 & 15  &  1  &  2  &  0  &  0  &  0  &  0  &  0  &  0  &  4  &  2  &  1  &  5  &  1  &  1  &  1  &  4  & 13  &  1.00  &  0  &  0  &  0  &  0  &  0  &  0  &  0  &  1  &  3  \\
  8849  &     5199.314103 & 15  &  1  &  2  &  0  &  0  &  0  &  0  &  0  &  0  &  4  &  4  &  1  &  5  &  1  &  1  &  1  &  4  & 14  &  1.00  &  0  &  0  &  0  &  0  &  0  &  0  &  2  &  1  &  1  \\
  8850  &     5387.416906 & 15  &  1  &  2  &  0  &  1  &  1  &  0  &  0  &  0  &  3  &  1  &  1  &  5  &  1  &  1  &  1  &  4  & 15  &  0.99  &  0  &  0  &  0  &  0  &  1  &  0  &  3  &  0  &  0  \\
  8851  &     5419.899376 & 15  &  1  &  2  &  0  &  1  &  1  &  0  &  0  &  0  &  3  &  3  &  3  &  5  &  1  &  1  &  1  &  4  & 16  &  0.98  &  0  &  0  &  0  &  0  &  1  &  0  &  0  &  2  &  1  \\
  8852  &     5467.412035 & 15  &  1  &  2  &  0  &  1  &  1  &  0  &  0  &  0  &  3  &  3  &  1  &  5  &  1  &  1  &  1  &  4  & 17  &  0.98  &  0  &  0  &  0  &  0  &  1  &  0  &  1  &  2  &  0  \\
  8853  &     5541.185628 & 15  &  1  &  2  &  1  &  0  &  0  &  0  &  0  &  0  &  2  &  2  &  1  &  5  &  1  &  1  &  1  &  4  & 18  &  1.00  &  0  &  0  &  1  &  0  &  0  &  0  &  0  &  1  &  1  \\
  8854  &     5596.854392 & 15  &  1  &  2  &  0  &  0  &  0  &  1  &  1  &  1  &  2  &  0  &  0  &  5  &  1  &  1  &  1  &  4  & 19  &  0.99  &  0  &  1  &  0  &  0  &  0  &  0  &  0  &  2  &  0  \\
  8855  &     5617.895727 & 15  &  1  &  2  &  0  &  0  &  0  &  1  &  1  &  1  &  2  &  2  &  2  &  5  &  1  &  1  &  1  &  4  & 20  &  0.95  &  0  &  1  &  0  &  0  &  0  &  0  &  0  &  0  &  2  \\
  8856  &     5650.480713 & 15  &  1  &  2  &  0  &  2  &  0  &  0  &  0  &  0  &  2  &  2  &  1  &  5  &  1  &  1  &  1  &  4  & 21  &  0.86  &  0  &  0  &  0  &  0  &  2  &  0  &  0  &  1  &  1  \\
  8857  &     5659.593538 & 15  &  1  &  2  &  0  &  0  &  0  &  1  &  1  &  1  &  2  &  2  &  1  &  5  &  1  &  1  &  1  &  4  & 22  &  0.85  &  0  &  1  &  0  &  0  &  0  &  0  &  0  &  1  &  1  \\
  8858  &     5683.390696 & 15  &  1  &  2  &  0  &  2  &  0  &  0  &  0  &  0  &  2  &  2  &  1  &  5  &  1  &  1  &  1  &  4  & 23  &  0.96  &  0  &  0  &  0  &  0  &  2  &  0  &  0  &  1  &  1  \\
  8859  &     5747.346056 & 15  &  1  &  2  &  1  &  1  &  1  &  0  &  0  &  0  &  1  &  1  &  1  &  5  &  1  &  1  &  1  &  4  & 24  &  1.00  &  0  &  0  &  1  &  0  &  1  &  0  &  1  &  0  &  0  \\
  8860  &     5829.908396 & 15  &  1  &  2  &  0  &  1  &  1  &  1  &  1  &  1  &  1  &  1  &  1  &  5  &  1  &  1  &  1  &  4  & 25  &  1.00  &  0  &  1  &  0  &  0  &  1  &  0  &  1  &  0  &  0  \\
  8861  &     5852.531380 & 15  &  1  &  2  &  1  &  0  &  0  &  1  &  1  &  1  &  0  &  0  &  0  &  5  &  1  &  1  &  1  &  4  & 26  &  0.95  &  0  &  0  &  2  &  0  &  0  &  0  &  0  &  0  &  0  \\
  8862  &     5862.518649 & 15  &  1  &  2  &  0  &  1  &  1  &  1  &  1  &  1  &  1  &  1  &  1  &  5  &  1  &  1  &  1  &  4  & 27  &  0.95  &  0  &  1  &  0  &  0  &  1  &  0  &  1  &  0  &  0  \\
  8863  &     5876.015235 & 15  &  1  &  2  &  0  &  3  &  1  &  0  &  0  &  0  &  1  &  1  &  1  &  5  &  1  &  1  &  1  &  4  & 28  &  0.92  &  0  &  0  &  0  &  0  &  3  &  0  &  1  &  0  &  0  \\
  8864  &     5901.578781 & 15  &  1  &  2  &  0  &  3  &  3  &  0  &  0  &  0  &  1  &  1  &  1  &  5  &  1  &  1  &  1  &  4  & 29  &  0.92  &  0  &  0  &  0  &  0  &  1  &  2  &  1  &  0  &  0  \\
  8865  &     6015.774283 & 15  &  1  &  2  &  0  &  0  &  0  &  2  &  2  &  1  &  0  &  0  &  0  &  5  &  1  &  1  &  1  &  4  & 30  &  1.00  &  1  &  1  &  0  &  0  &  0  &  0  &  0  &  0  &  0  \\
  8866  &     6064.380191 & 15  &  1  &  2  &  0  &  2  &  0  &  1  &  1  &  1  &  0  &  0  &  0  &  5  &  1  &  1  &  1  &  4  & 31  &  1.00  &  0  &  1  &  0  &  0  &  2  &  0  &  0  &  0  &  0  \\
\hline
\hline
\end{tabular}
\begin{tabular}{cll}
             Column       &    Notation                 &      \\
\hline
   1 &   $N$              &       Level number (row)    \\
   2 & $\tilde{E}$        &       Term value (in \cm)                           \\
   3 &  $g_{\rm tot}$     &       Total degeneracy   \\
   4 &  $J$               &       Rotational quantum number    \\
   5 &  $\Gamma_{\rm tot}$&       Total symmetry in \Td(M)                     \\
   6,7,9,12 &  $n_1-n_4$  &       Normal mode vibrational quantum numbers       \\
   8,10,13 &  $L_2, L_3, L_4$ &   Vibrational angular momenta quantum  numbers\\
   11,14 &  $M_3, M_4$    &       Vibrational angular momenta quantum  numbers    \\
  15 &  $\Gamma_{\rm vib}$&       Symmetry of the vibrational contribution in \Td(M) \\
  16 &  $J$               &       Rotational quantum number (the same as column 2)                             \\
  17 &  $K$               &       Rotational quantum number, projection of $J$ onto the $z$-axis                \\
  18 &  $\tau_{\rm rot}$  &       Rotational parity (0,1)                                                        \\
  19 &  $\Gamma_{\rm rot}$& Symmetry of the rotational contribution in \Td(M) \\
  20 &  $N_{ J,\Gamma}$   &       Running number in the $J,\Gamma$ block  \\
  21 &  $|C_i|^2$         &       Largest coefficient used in the assignment \\
  22-30 &  $v_1-v_9$         &       Local mode vibrational quantum numbers (see \protect\citep{jt555}) \\
\hline
\end{tabular}
\end{table}

\begin{table}
\caption{\label{t:Transit-file} Extract from the 10to10 Transition file.}
\begin{center}
\renewcommand{\arraystretch}{1.0}
\begin{tabular}{rrc}
    \hline
    \hline
         $F$  &  $I$  & A$_{\rm IF}$ / s$^{-1}$ \\
\hline
           1002348 &     1180308& 1.2167e-04 \\
           1033584 &     1024255& 4.5595e-04 \\
             10461 &       27572& 1.2159e-03 \\
           1046761 &     1199185& 1.5956e-03 \\
           1049153 &      863924& 4.4509e-04 \\
           1049953 &     1024823& 4.1242e-04 \\
           1050761 &     1024875& 7.6355e-05 \\
            105546 &      135097& 1.0140e-03 \\
           1088309 &     1292264& 2.6590e-06 \\
           1099011 &     1111107& 7.0156e-06 \\
           1116545 &     1058555& 4.6582e-04 \\
           1150480 &     1236818& 1.5768e-04 \\
    \hline
    \hline
\end{tabular}

\noindent
 $I$: Upper state counting number;
$F$:      Lower state counting number;
$A_{IF}$:  Einstein A coefficient in s$^{-1}$.

\end{center}

\end{table}

\section{Analysis}
\label{s:analysis}

In this section we compare the $^{12}$CH\4\ spectra generated using
the 10to10 line list with data available from other sources.

\subsection{HITRAN}
\label{s:HITRAN}

\begin{figure}
\caption{Absorption of $^{12}$CH\4\ at $T=296$~K: HITRAN~2012 (bottom) \protect\citet{jt557} and 10to10 (top). The yellow stars are to indicate the intensity coverage in HITRAN.}
\centering
{\leavevmode \epsfxsize=11.0cm \epsfbox{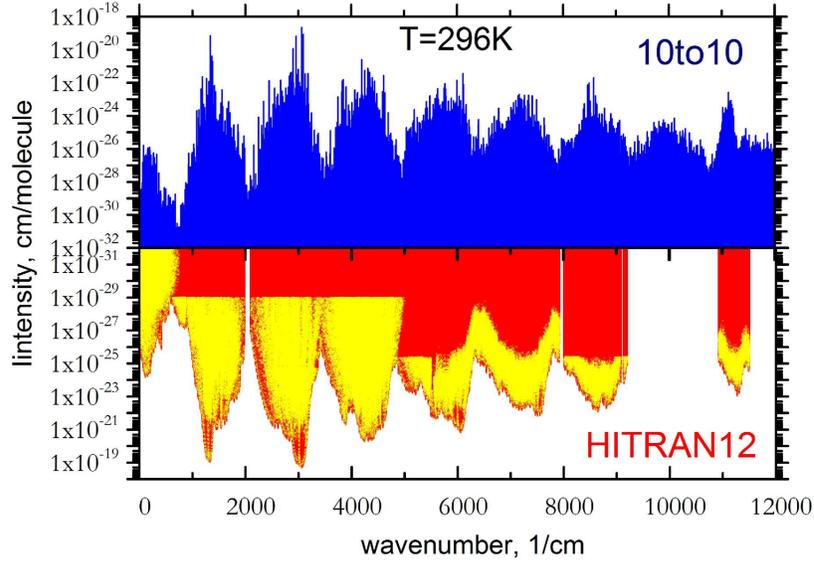}
\label{f:hitran}}
\end{figure}

The latest (2012) update of HITRAN contains 336\,830 lines
of $^{12}$CH\4 \citep{jt557}. Figure~\ref{f:hitran} compares a $T=296$~K HITRAN
spectrum for $^{12}$CH\4\ with the spectrum generated from 10to10 at
the same temperature. The 10to10 spectral lines are given as sticks
with the intensity (cm/molecule) represented by their height. In order
to simplify the plot of the 10to10 spectra, only the strongest lines
within each 0.01 bin are shown. The HITRAN~2012 data are given as
circles. Apart from general good agreement between theory and
experiment, one can see that the coverage of the experimental CH\4\
data as represented by HITRAN though reasonably good, still has gaps
in both the frequency and intensity directions.

\begin{figure}
\caption{ The absorption spectrum of $^{12}$CH\4\ at $T=296$~K: 10to10 (blue) vs HITRAN~2012 (red) compared for different wavenumber windows. The bottom-right display  also shows the 296~K spectrum convolved with the doppler profile (green).
}
\centering
{\leavevmode \epsfxsize=16.0cm \epsfbox{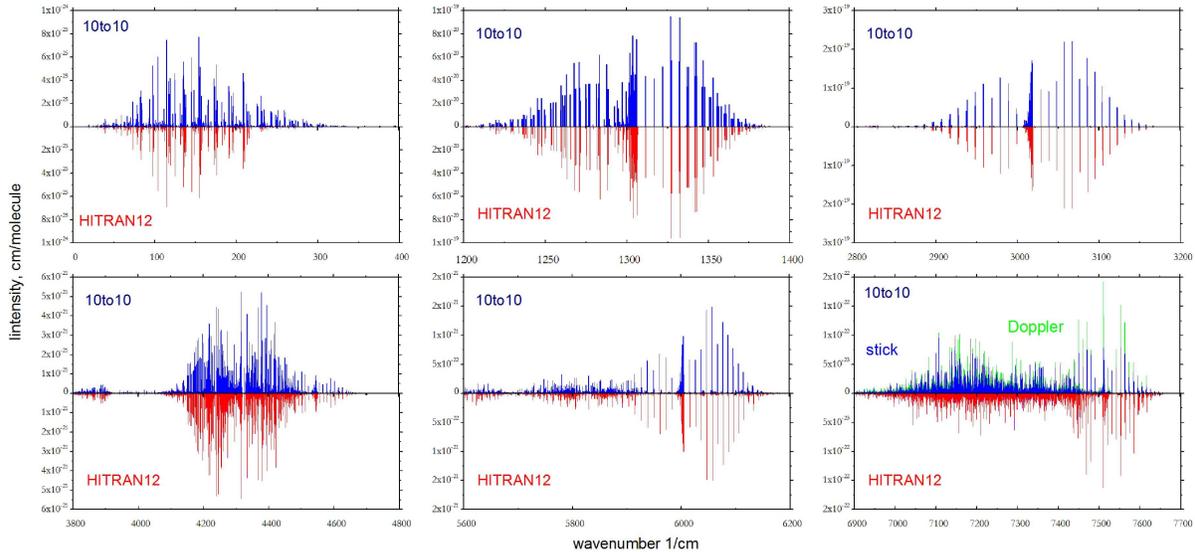}
\label{f:hitran:overview}}
\end{figure}

Figure~\ref{f:hitran:overview} gives a number of zoomed-in pictures of
several bands, which illustrate the quality of our data. Both the line positions and the absolute
intensities agree well with the HITRAN data, at least for the
strongest lines. A more detailed line-by-line comparative analysis is
planned in the nearest future.  Our spectrum seems to disagree with
HITRAN~2012 in the wavenumber region 800--900~\cm, where new release
of HITRAN has a stronger feature which was not present in the
previous, 2008 release of HITRAN \citep{jt453} or in our calculations.
This feature corresponds to two hot, combination bands $3\nu_4-2\nu_2$ and
$3\nu_4-\nu_3$.  These bands are yet to be characterized experimentally and
thus result purely from an extrapolation.  We suggest that there is a
problem with the effective dipole moment model used to generate these
lines.  Furthermore, we note that the new HITRAN~2012 lines in the
region 300--600~\cm\ have significantly different (lower)
intensities compared to the data in HITRAN~2008 where the latter also
agrees well with the 10to10 spectrum.

The intensities of our stick spectrum in the band around the 7000~\cm\
band appear to be very different from those of HITRAN.  The
bottom-right display of Fig.~\ref{f:hitran:overview} also shows the
spectrum convolved with a $T=296$~K Doppler profile, which allows
overlapping lines to be added together. This spectrum is in much
better agrement with the HITRAN stick intensities, suggesting that
most of the  intensities given as single lines in HITRAN are actually from blends.
Our stick spectrum
starts `de-focusing' at 8000~\cm\ in terms of the centre of the bands,
this probably indicates the limitations of our refined PES in this
region.

\subsection{Hot spectrum of CH\4}
\label{s:hot}

\subsubsection{Experimental comparisons}

\begin{figure}
\caption{Temperature dependence of the absorption  cross sections (cm/molecule) of $^{12}$CH\4\ computed at $T=300$, 1000, 1500, and 2000~K using the 10to10 line list.  }
\centering
{\leavevmode \epsfxsize=11.0cm \epsfbox{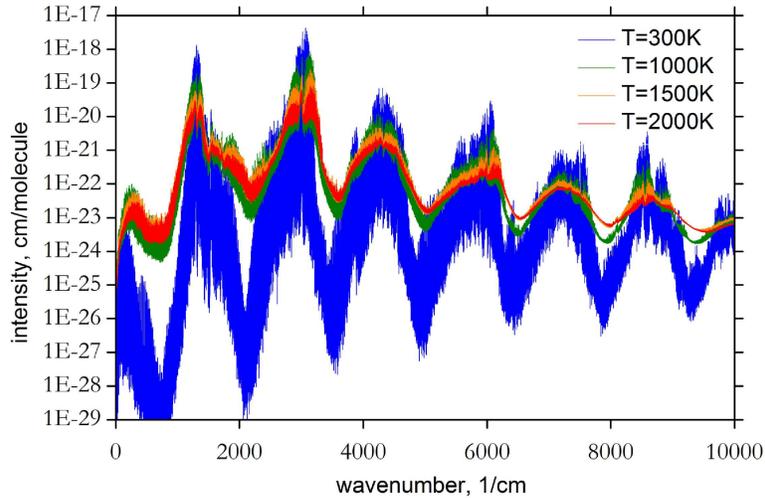}
\label{f:temp}}
\end{figure}

It is common for the absorption spectra of polyatomic molecules to
change dramatically with increased temperature.  Figure~\ref{f:temp}
illustrates the dynamic temperature effect in case of the methane
absorption spectrum.  Especially the window regions between strong
bands gain significant intensity as temperature increases. These
windows regions are generally under-sampled in laboratory experiments because
of their very weak absorption. The corresponding lines compiled mostly
from the hot-bands transitions become increasingly important at elevated
temperatures when the hot-bands get stronger.
The HITRAN-based spectrum
significantly underestimates the opacity of CH\4\ in the regions
between the strong bands and, as a result, gives the bands the wrong shape.
This is especially evident at the higher
wavenumbers bands, 5000-6000~\cm\ and 7000-8000~\cm, where the HITRAN
representation of methane is especially sparse. To illustrate the
problem of the under-sampling, Table~\ref{t:colors} lists opacities
(integrated intensities) for the $H$, $J$ and $K$  spectral regions (see also
below). We used a simple definition of the $K$ (2.0-2.4~$\mu$m), $H$ (1.5-1.8~$\mu$m) and $J$ (1.1-1.4~$\mu$m) regions given also in Table~\ref{t:colors}.

\begin{table}
\caption{ Summary of available experimental and theoretical methane line lists. }
\label{t:statistics}
\footnotesize
\begin{center}
\begin{tabular}{lcccrr}
  \hline
  \hline
Line list                          &             $T$, K            &     wavenumber, \cm     &    $N$ lines           & $J_{\rm max}$& $E_{\rm max}$ \\
  \hline
\citet{03NaBexx.CH4}               &        800, 1\,000, 1\,273        &        2\,000-6\,400        &    16\,191,23\,002,26\,821  &              &               \\
\citet{12HaBeMi.CH4}               &     573, 673, \ldots, 1\,673    &         960-5\,000        &    33\,955-79\,098         &              &               \\
\citet{08ThGeCa.CH4}               &     1\,005, 1\,365, 1\,485, 1\,625    &        2\,700-3\,300        &4\,904, 3\,824, 5\,647, 2\,381, 1\,559 &              &               \\
\\
HITRAN 2012                        &                               &         0-11\,502         &               336\,830   &         25   &        4\,263   \\
\citet{09WaScSh.CH4}               &                               &          0-6\,133         &              1\,410\,789   &         34   &        6\,200   \\
MeCaSDa                            &                               &          0-6\,446         &              5\,045\,392   &         25   &        6\,340   \\
10to10                             &                               &         0-12\,000         &           9\,819\,605\,160   &         39   &        8\,000   \\
  \hline
\end{tabular}
\end{center}
\end{table}

\begin{table}
\caption{ Integrated intensities for the $K$, $H$, and $J$ bands from different line lists.  }
\label{t:colors}
\footnotesize
\begin{center}
\begin{tabular}{lcccccc}
  \hline
  \hline
                      &  296~K   &  1000~K  &  1200~K  &  1273~K  &  1500~K  &  2000~K    \\
  \hline
  Line list           &  \multicolumn{6}{c}{$K$ band: 4166.7 -- 5000.0 \cm}              \\
  \hline
\citet{03NaBexx.CH4}  &          &  2.6E-19 &          &  1.6E-19 &          &            \\
\citet{12HaBeMi.CH4}$^a$  &          &  5.2E-19 &          &  3.7E-19 &          &            \\
\citet{08ThGeCa.CH4}  &          &          &          &          &          &            \\
HITRAN 2012           &  8.1E-19 &  1.6E-19 &  2.0E-19 &  1.7E-19 &  9.7E-20 &   3.0E-20  \\
\citet{09WaScSh.CH4}  &  8.1E-19 &  3.5E-19 &  4.6E-19 &  3.8E-19 &  2.3E-19 &   7.1E-20  \\
MeCaSDa               &  8.1E-19 &  2.8E-19 &  4.0E-19 &  3.4E-19 &  2.2E-19 &   7.6E-20  \\
10to10                &  7.8E-19 &  8.0E-19 &  8.2E-19 &  8.2E-19 &  7.8E-19 &   5.7E-19  \\
  \hline
                      &  \multicolumn{6}{c}{$H$ band: 5555.6 -- 6666.7~\cm}                                     \\
\citet{03NaBexx.CH4}  &          &          &          &  2.3E-20 &          &            \\
\citet{12HaBeMi.CH4}  &          &          &          &          &          &            \\
\citet{08ThGeCa.CH4}  &          &          &          &          &          &            \\
HITRAN 2012           &  1.0E-19 &  1.0E-20 &  1.1E-20 &  8.9E-21 &  4.7E-21 &   1.2E-21  \\
\citet{09WaScSh.CH4}  &  2.3E-19 &  1.5E-20 &  1.6E-20 &  1.2E-20 &  6.3E-21 &   1.6E-21  \\
MeCaSDa               &  1.3E-19 &  2.5E-20 &  3.0E-20 &  2.4E-20 &  1.3E-20 &   3.8E-21  \\
10to10                &  1.4E-19 &  1.5E-19 &  1.6E-19 &  1.6E-19 &  1.5E-19 &   1.1E-19  \\
  \hline
                      &  \multicolumn{6}{c}{$J$ band: 7142.9 -- 9090.9~\cm}                                     \\
\citet{03NaBexx.CH4}  &          &          &          &          &          &            \\
\citet{12HaBeMi.CH4}  &          &          &          &          &          &            \\
\citet{08ThGeCa.CH4}  &          &          &          &          &          &            \\
HITRAN 2012           &  1.4E-21 &  7.3E-23 &          &          &  3.0E-23 &   7.3E-24  \\
\citet{09WaScSh.CH4}  &          &          &          &          &          &            \\
MeCaSDa               &          &          &          &          &          &            \\
10to10                &  6.0E-20 &  6.3E-20 &          &          &  6.6E-20 &   5.2E-20  \\
  \hline
\end{tabular}

$^a$  The actual temperature used by \citet{12HaBeMi.CH4} was $T=973$~K.

\end{center}
\end{table}

Experimentally, the spectrum of hot methane has been the subject of several
studies, see \citet{03NaBexx.CH4} ($T=$ 800, 1\,000, and 1273~K),
\citet{08ThGeCa.CH4} ($T=1005- 1\,820$~K), and \citet{12HaBeMi.CH4} ($T=573-1\,673$~K)~\cm.
Table~\ref{t:statistics} summarises the coverage
of these studies.  The 10to10 results given in this table suggest that the absorption by methane
in each band is essentially independent of temperature; all previous
studies show significantly less absorption by methane at elevated temperatures.

\begin{figure}
\caption{ The absorption spectrum of $^{12}$CH\4\ at $T=1485$~K.
The bottom display shows the experimental intensities (cm/molecule) derived by
\protect\citet{08ThGeCa.CH4} from their emission spectra in the region 2700--3300~\cm. The top display shows: (sticks) the 10to10 absorption spectrum lines obtained as the maximal intensity at the 0.2~\cm\ wavenumber bin  and (Doppler) the 10to10 absorption spectrum convolved with the Doppler profile  and integrated over a 0.1~\cm\ wavenumber bin. }
\centering
{\leavevmode \epsfxsize=11.0cm \epsfbox{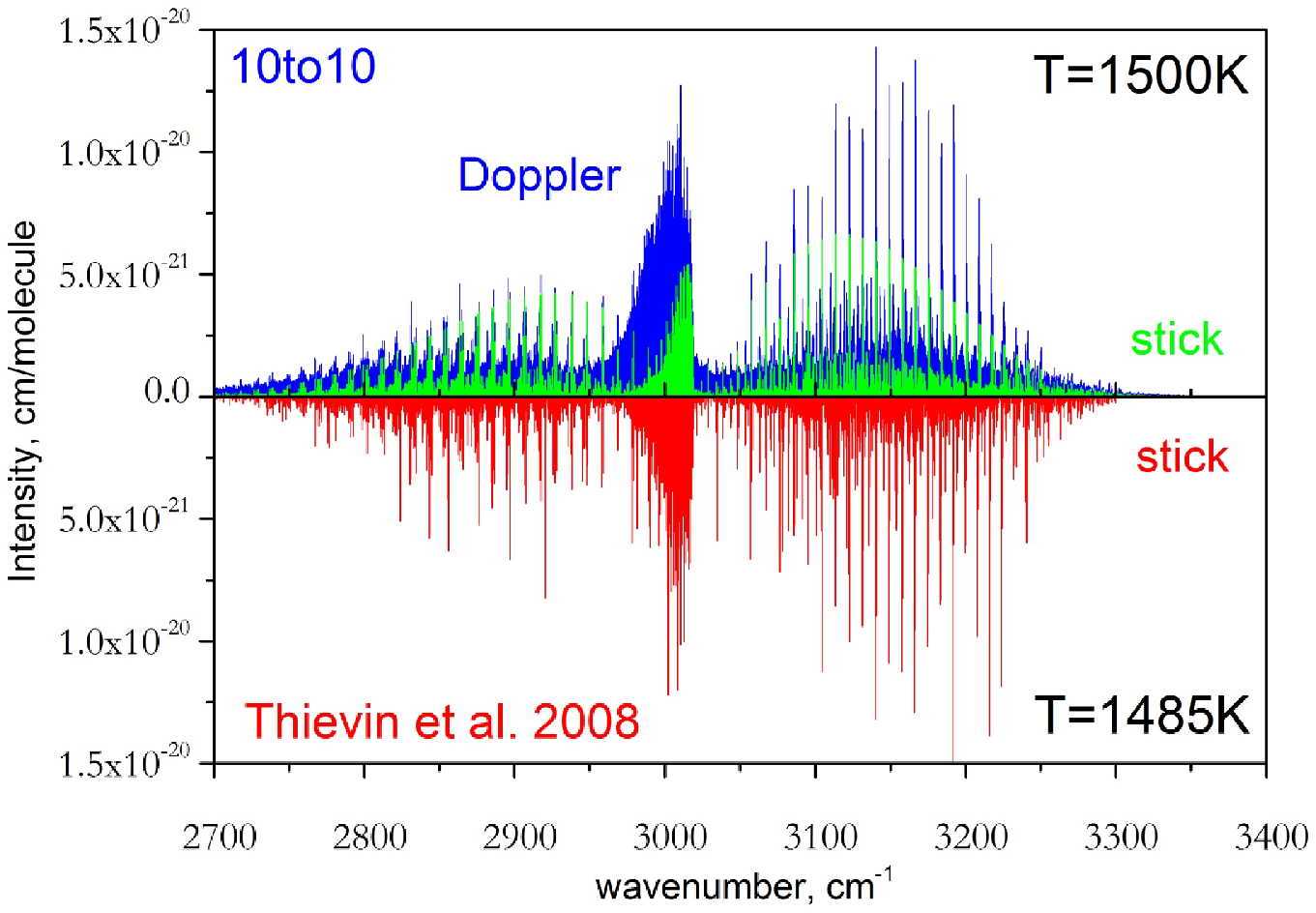}
\label{f:thievin}}
\end{figure}

Figure~\ref{f:thievin} shows the $T=1\,425$~K absorption spectrum of
\citet{08ThGeCa.CH4} compared to two types of synthetic (10to10) absorption spectra of CH\4\ at
$T=1\,500$~K: (a) a `stick' spectrum and (b) a Doppler-broadened spectrum. The good agreement with the latter spectrum shows the importance of the correct description of the line mixing for accurate reconstruction of the line absorption, absent in the experimental line list by \citet{08ThGeCa.CH4}.

\begin{figure}
\caption{Experimental (crosses) and theoretical (blue bars) absorption intensities   of $^{12}$CH\4\ at $T=1000$~K and $1500$~K. The experimental data are from \protect\citet{03NaBexx.CH4} (red) and  \protect\citet{12HaBeMi.CH4} (green). Note that the actual temperatures used by \citet{12HaBeMi.CH4} were $T=973$~K and $1473$~K.
}
\centering
{\leavevmode \epsfxsize=11.0cm \epsfbox{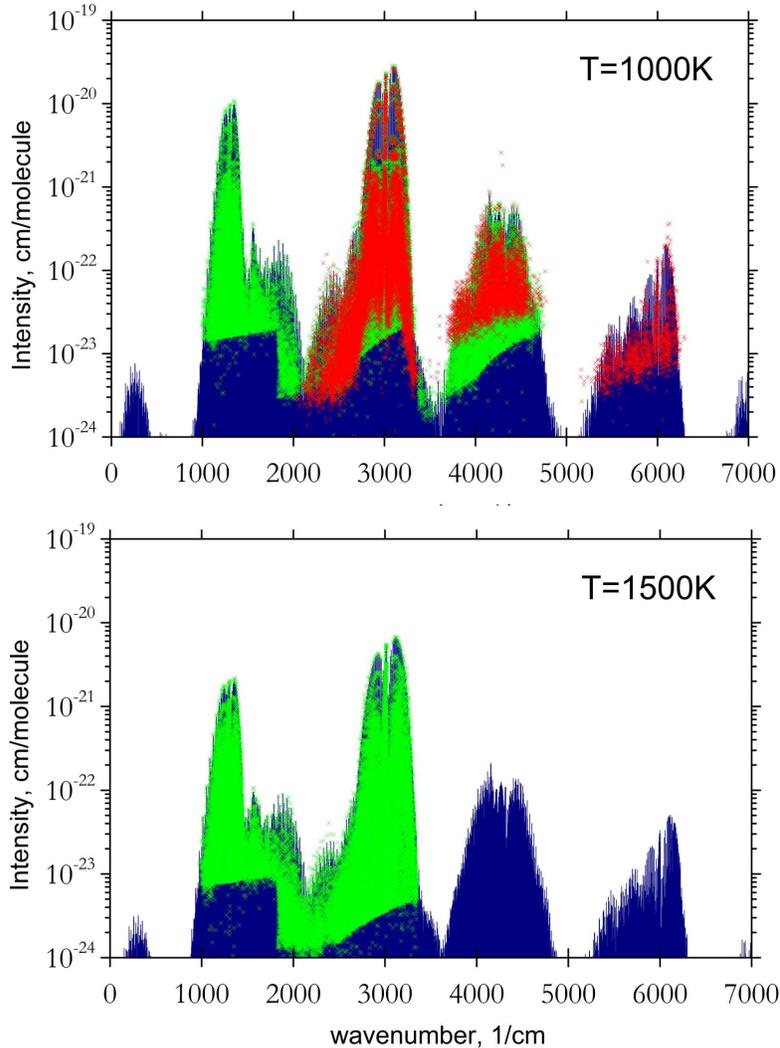}
\label{f:bernath}}
\end{figure}

Figure~\ref{f:bernath} compares the
10to10 absorption (stick) spectrum of CH\4\ at $T=1000$~K with the
 1000~K absorption spectrum of \citet{03NaBexx.CH4}  and to the $T=973$~K spectrum of
\citet{12HaBeMi.CH4}. We see that there is generally
good agreement between theory and experiment. However the integrated
absorption intensities are very different, see Table~\ref{t:colors}.
In the region 4166.7 -- 5000.0~\cm\ ($K$-band) at $T=1273$~K the line
list of \citet{12HaBeMi.CH4} gives $I_{\rm tot} = $ 3.7$\times
10^{-19}$~cm/molecule, compared to our value, 8.2$\times
10^{-19}$~cm/molecule. HITRAN~2012
underestimates the opacity at these temperature and region even more,
giving $I_{\rm tot} = $ 1.7$\times 10^{-19}$~cm/molecule.  For the higher
wavenumber regions ($H$ and $J$) the problem of missing opacities is
even more pronounced from one \citep{03NaBexx.CH4} to two (HITRAN
2012) orders of magnitude.

\begin{figure}
\caption{Comparison of the 10to10 and \protect\citet{PNNL} cross sections of $^{12}$CH\4\ at $T=50$~C.   }
\centering
{\leavevmode \epsfxsize=11.0cm \epsfbox{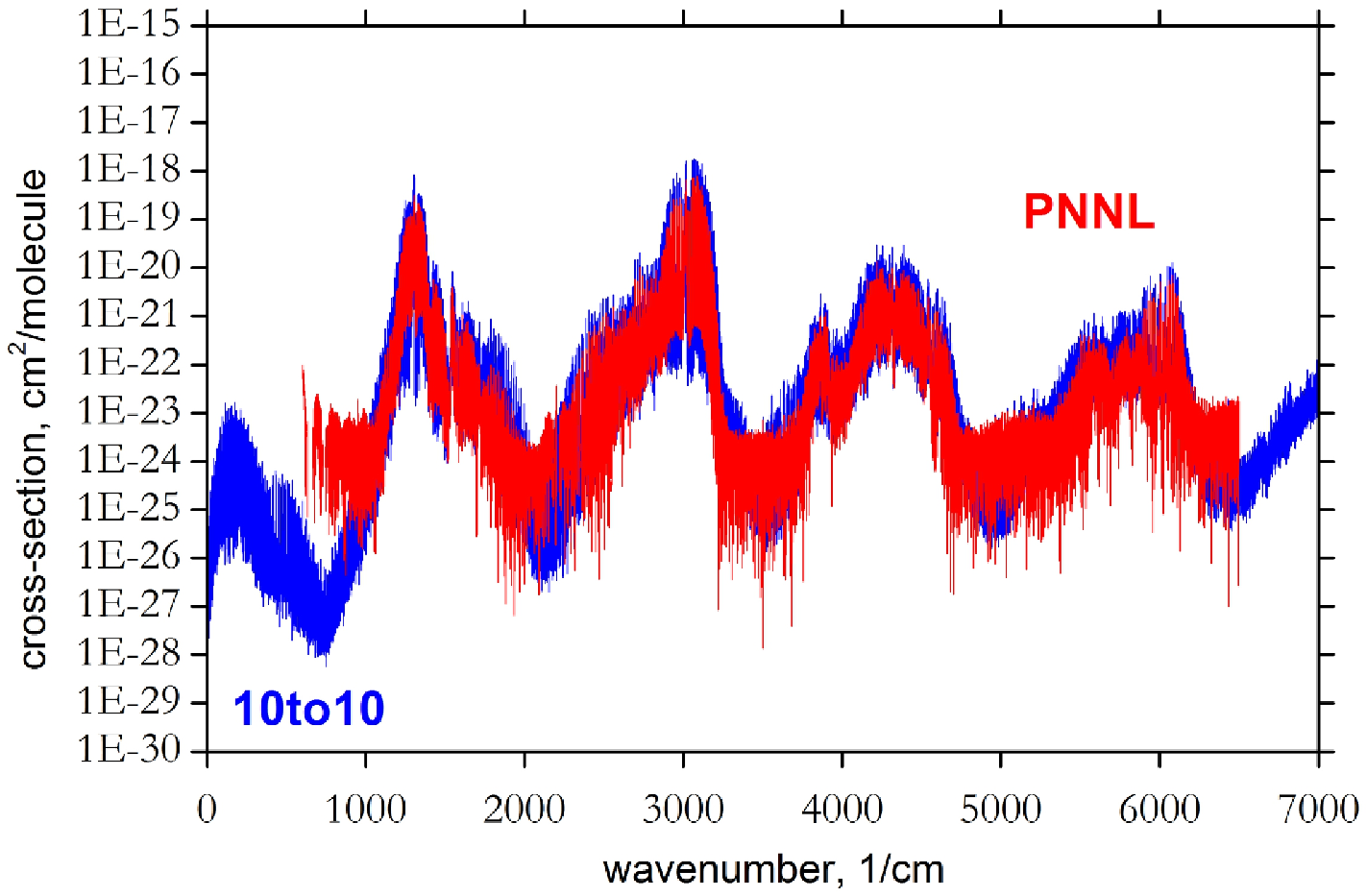}
\label{f:PNNL}}
\end{figure}

A final piece of experimental  information are the ``PNNL'' cross sections
provided by \citet{PNNL} (cm$^2$/molecule) which are presented at three
temperatures, $T$=5, 20, and 50~C.
Fig.~\ref{f:PNNL} compares the $T=50$~C PNNL cross
sections with that generated by 10to10.

\subsubsection{Theoretical comparisons}

\begin{figure}
\caption{Theoretical cross sections $\sigma$ (cm$^2$/molecule) of $^{12}$CH\4\ at  $T=1500$~K obtained using three line lists: (i) 10to10, (ii) MeCaSDa \protect\citep{13BaWeSu.CH4}, and (iii) \protect\citet{09WaScSh.CH4}.}
\centering
{\leavevmode \epsfxsize=11.0cm \epsfbox{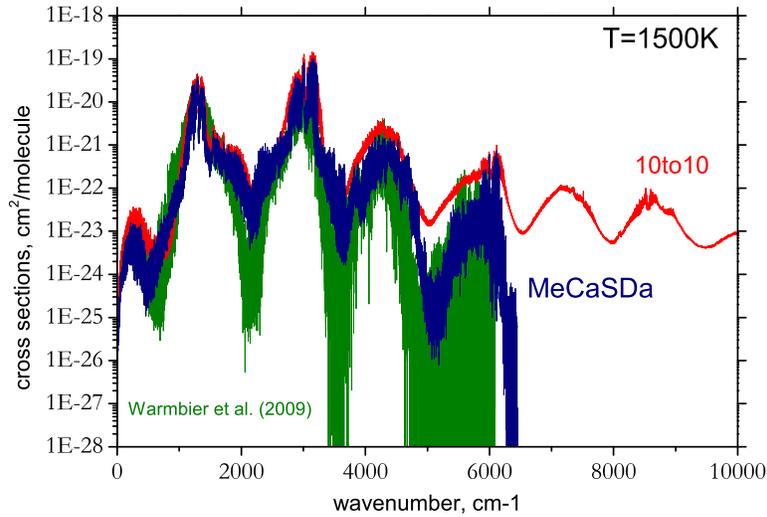}
\label{f:mecasda}}
\end{figure}

\citet{09WaScSh.CH4} reported a theoretical hot-line list for
methane computed variationally using the program MULTIMODE \citep{98CaBoxx.method} in
conjunction with an \ai\  PES and DMS calculated at the
RCCSD(T)/aug-cc-pVTZ level of theory. This line list contains 1.4 million lines
covering the wavenumbers  up to 6200~\cm\ with $J = 0\ldots 34$
and energies up to 6200~\cm\ only. We used this line list to generate
the cross sections of CH\4\ at $T=1500$~K assuming Doppler
broadening only. Fig.~\ref{f:mecasda} compares these cross sections
with those from  10to10 at the same
temperature. The significantly lower values of the cross sections by
\citet{09WaScSh.CH4} indicate that their line list is very incomplete
especially for higher frequencies and for hot bands.  The corresponding
integrated intensities for the $H$ and $K$ bands listed in Table~\ref{t:colors}
give a quantitative illustration of this conclusion.  For example, the
$H$-band integrated intensity at $T=1\,500$~K obtained from the line list
of \citet{09WaScSh.CH4} is
$6.3\times 10^{-21}$~cm$/$molecule which is ten time weaker than the 10to10 value.

Very recently an accurate synthetic line list MeCaSDa
was reported by \citet{13BaWeSu.CH4}. It was generated
using the XTDS software \cite{08WeBoRo.method}. This line list contains 5~045~392 transitions of $^{12}$CH\4\
with $J \le 25$ and covering the wavenumber range up to 6446~\cm\ with lower state energies up to
6340~\cm, see Table~\ref{t:statistics}. The line positions are
close to experimental accuracy with a reasonable coverage
of CH\4\ hot transitions.  Figure~\ref{f:mecasda} compares the
$T=1\,000$~K spectrum generated using MeCaSDa (below $6\,446$~\cm)
and that obtained with 10to10. The differences are less pronounced
than for  HITRAN or \citet{09WaScSh.CH4}, however some opacity at
high temperature is still missing, see Table~\ref{t:colors}. For
example, for the $H$-band integrated intensity at $T=1\,000$~K MeCaSDa gives
$I_{\rm tot} = $ 2.5$\times 10^{-20}$~cm/molecule, which is 6 times
smaller than that of 10to10. At $T=1\,500$~K this difference is even
more dramatic, 1.3$\times 10^{-20}$ (MeCaSDa) compared to 1.5$\times
10^{-19}$~cm/molecule (10to10).

\subsubsection{Astrophysical comparisons}

\begin{figure}
\caption{Jupiter spectrum  recorded 17 July 1994 during the impact of Comet Shoemaker-Levy 9 \protect\citep{jt198} (red) and 10to10 emission spectrum of CH\4\ at $T=1400$~K. }
\centering
{\leavevmode \epsfxsize=11.0cm \epsfbox{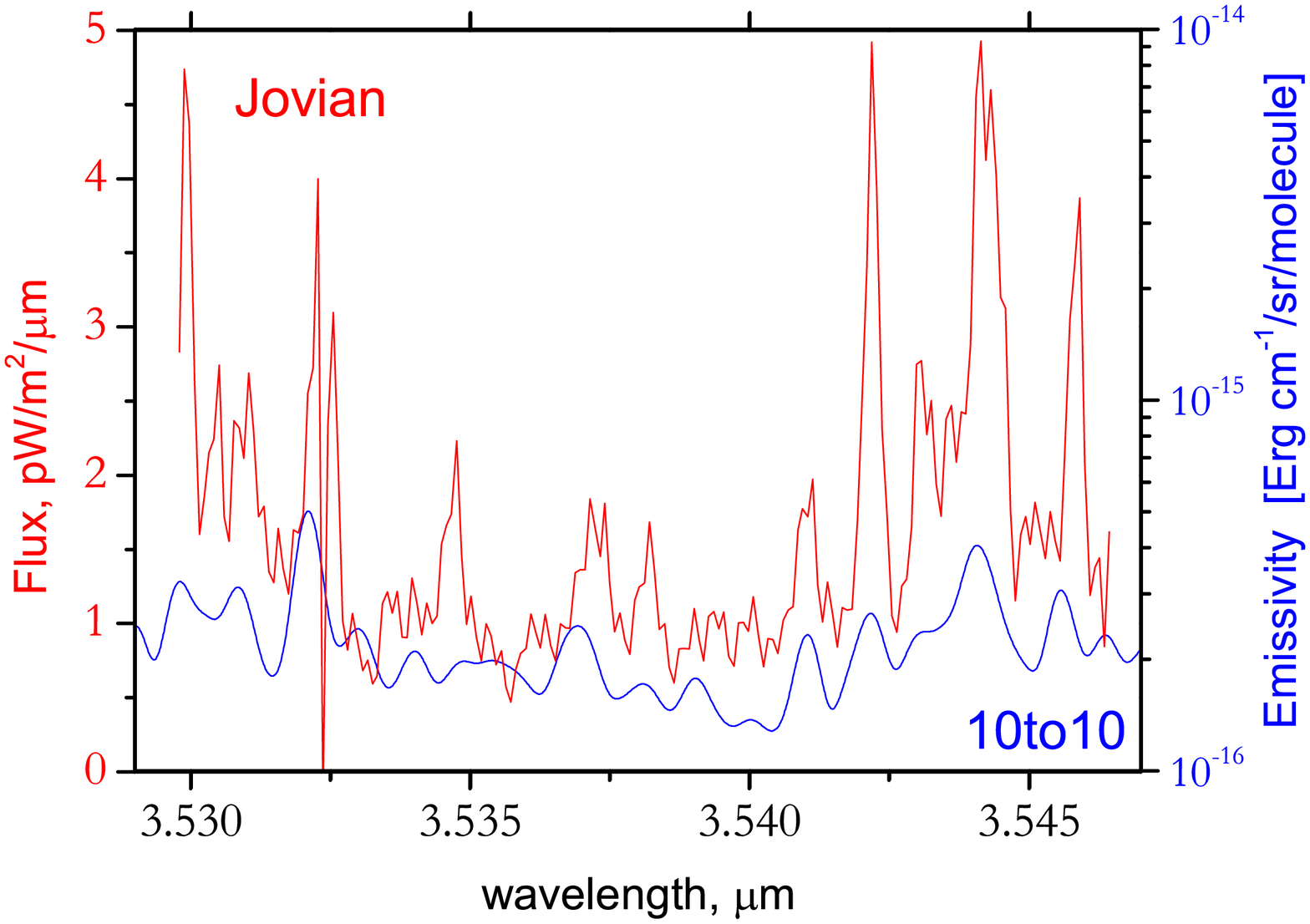}
\label{f:comet}}
\end{figure}

Methane has been detected in the atmospheres of exoplanets,
cool stars and comets. We have selected two extra-terrestrial
methane spectra recorded at elevated temperatures at relatively high or medium
resolution. The several collisions of  Comet Shoemaker-Levy 9 with Jupiter
was carefully observed in the infrared \citep{jt154}. Fig.~\ref{f:comet} compares a synthetic 10to10 emission
spectrum of methane at $T=1400$~K with one of the Jovian spectra
recorded during the impact of Comet Shoemaker-Levy 9 on 17 July 1994
\citet{jt198}. \citet{jt198} identified a number of hot methane features
based on the high resolution spectrum of \citet{94HiChTo.CH4}.

\begin{figure}
\caption{ The spectrum of the T5 Brown Dwarf 2MASS J0559-1404 \protect\citep{05CuRaWi.dwarfs} compared to the 10to10 absorption spectrum of CH\4\ at $T=1\,200$~K.
All features are from CH\4. Arrows are shown to guide the eye.}
\centering
{\leavevmode \epsfxsize=11.0cm \epsfbox{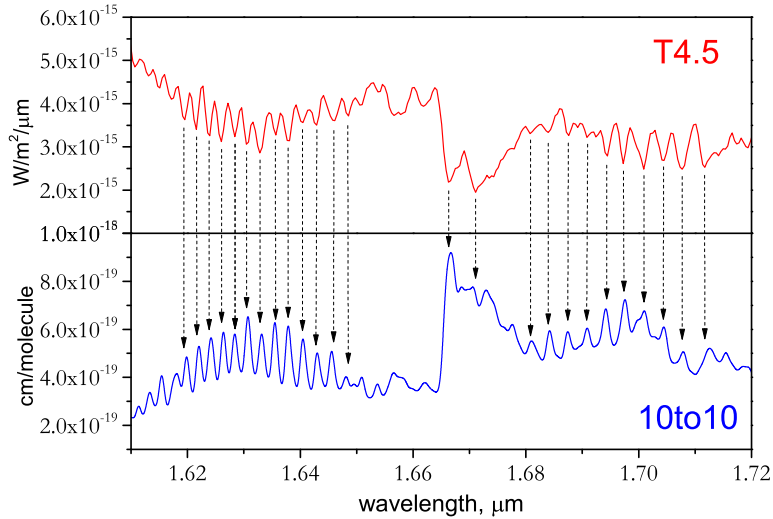}
\label{f:dwarf}}
\end{figure}

Our other example is associated with the spectroscopy of the methane brown dwarfs.
Fig.~\ref{f:dwarf} gives a comparison of a synthetic 10to10 absorption spectrum of CH\4\
at $T=1\,200$~K with the spectrum of the T4.5  brown dwarf 2MASS J0559-1404 centered at
1.67~$\mu$m  reported by \citet{05CuRaWi.dwarfs}, who characterised the methane spectral
features based on the high resolution CH\4\
spectrum by \citet{03NaBexx.CH4}. In both cases our spectra agree well with
the observation and can potentially offer a more complete picture of
the methane contributions to the spectral description of cool stars
and exoplanets.

\section{Conclusion}
\label{s:conclusion}

We have computed a methane vibration-rotation line list containing almost 10 billion transitions
which we call 10to10. The full
line list for each of these isotopologues can be downloaded from the CDS,
via \url{ftp://cdsarc.u-strasbg.fr/pub/cats/J/MNRAS/} or \url{http://cdsarc.u-strasbg.fr/viz-bin/qcat?J/MNRAS/}, and from
the ExoMol website \url{www.exomol.com}.
The problem with having such an extensive line list is that files are large and difficult to manipulate.
The ExoMol project provides facilities for generating cross sections at appropriate temperatures
\citep{jt542}.

10to10 has been constructed as the next member
in the series of line lists provided by the ExoMol
project \citep{jt528,jt529,jt563,jt570} which aims to provide comprehensive line lists for studies of
hot atmospheres such as exoplanet, brown dwarfs and cool stars. It
construction means that there are now hot line lists for the key set
of molecules methane, water \citep{jt378} and ammonia \citep{jt500}.
These provide a suitable set of spectroscopic data on hot molecules for
systematic models to be performed on astronomical objects with hot,
molecular hydrogen-rich atmospheres. Work in this direction has already
started.

However, despite the size of the 10to10 line list it remains
incomplete and should only be used with caution for temperatures
higher than 1500 K as it will miss a significant proportion of the
opacity at these higher temperatures. Furthermore, the effective,
empirical character of the underlying potential energy surface and the
convergence deficiency of the TROVE nuclear motion calculations means
that the accuracy of the energy levels, particularly at higher
energies, and hence the related transition frequencies could be
improved.  Work in this direction will be undertaken in due course.



\section*{Acknowledgements}

This work was supported by STFC and ERC Advanced Investigator Project 267219. The research
made use of the DiRAC@Darwin and DiRAC@COSMOS HPC clusters. DiRAC is the UK HPC facility for
particle physics, astrophysics and cosmology and is supported by STFC and BIS.
We also thank UCL for use of the Legion High Performance Computer for
performing the electronic structure calculations and Bianca Maria Dinelli for providing her cometary data.


\section*{A Appendix: Methane quantum numbers}

Methane is a symmetric five atomic molecule characterised by nine
vibrational degrees of freedom with a vanishing permanent dipole
moment.  It is a very high symmetry molecule of the \Td(M) symmetry
group.  In this work we use the Molecular Symmetry group
\citep{04BuJexx.method} to describe the irreducible representations
(irreps) of CH\4. Thus the ro-vibrational states of CH\4\ will be
assigned with the five irreps, $A_1$, $A_2$, $E$, $F_1$, and $F_2$,
where $A_1$ and $A_2$ are one-dimensional (1D), $E$ is a 2D, and $F_1$
and $F_2$ are 3D irreps. Because of the $i=\frac{1}{2}$ nuclear spins of the
hydrogens, the total spin-rotation-vibration states of $^{12}$CH\4\ is
fermionic and degenerate if
hyperfine splittings are neglected. This degeneracy gives rise to nuclear
statistical weight factors of 5, 5, 2, 3, and 3 for $\Gamma$ = $A_1$,
$A_2$, $E$, $F_1$, and $F_2$, respectively.  The symmetry $\Gamma$ is
a `good quantum number' \citep{04BuJexx.method} as well as the
rotational angular momentum quantum number $J$, which are two main
labels used to classify the ro-vibrational states of CH\4. The
electric dipole selection rules are given in
Eqs.~(\ref{e:selection-rules}) and (\ref{e:selection-rules:J}).  The dipole
moment components of the molecule in the body-fixed frame span the
three components of the $F_{2}$  irrep, $F_{2_x}$, $F_{2_y}$,
and $F_{2_y}$ and the potential energy function is fully symmetric
($A_1$).

Apart from these well-defined, good quantum numbers, like $J$ and
symmetry, it is a common practise in spectroscopy to assign the
ro-vibrational states of a molecule using a full set of rotational and
vibrational labels which aid characterizing the state in question.
These quantum numbers are usually associated with some selected
referenced ro-vibrational functions and are used to describe the
similarity of the given ro-vibrational eigenfunction to these
reference functions.  In variational approaches, where the
eigenfunction is given as an linear expansion in terms of the basis
functions, the similarity can be assessed by identifying the largest
contribution to such an expansion (see, for example
\citep{07YuThJe.method}). In this case the basis functions play the
role of the reference functions and the labels characterising the
corresponding basis functions are used as the approximate quantum
numbers for the eigenstate. The standard problem with this, and other
approaches, is the strong mixing of basis set functions at high
excitations which gives rise to the ambiguity in assignment.
Alternatively, the similarity can be established by computing overlaps
between the target and reference wavefunctions, see, for example
\citet{10MaFaSz.method}, where the reference functions can be any
functions and are not limited by the choice of the basis set. The
conventional choice of the reference functions includes the
rigid-rotor functions for the rotation and the normal mode
(degenerate) harmonic oscillator eigenfunctions. Even the latter
approach suffers from the state mixing problem and thus cannot
guarantee unambiguous definition of the quantum numbers.

The difficulty of assigning the methane energy levels is well
recognized. \citet{06BoReLo.CH4} suggested the following quantum
numbers for the vibrational assignment: $n_i$, $l_i$, $m_i$ and $C_i$,
$i=3,4$, where  $n_i$ is the vibrational quantum number, $l_i$ is the
vibrational angular momentum ($l_i = n_i, n_i - 2, n_i -4,\ldots ,0$
or $1$ for $i=2,3,4$), $C_i$ is a \Td\ irrep and $m_i$ is multiplicity
index for a given  set $(n_i,l_i,C_i)$. We partly follow this
scheme.

We use the largest contribution approach to assign the eigenfunction
of methane. Our selection of the quantum numbers for CH\4\ is given in
Eq.~\eqref{e:qns}. For the rotation basis, symmetrized rigid-rotor
functions $|J,K,\Gamma_{\rm rot}\rangle$ are used, where $K = |k|$,
$k$ is the projection of $\mathbf J$ on the body-fixed $z$-axis, and
$\Gamma_{\rm rot}$ is the rotational symmetry. For the vibrations we
use nine 1D local (non-normal) mode basis functions
$\phi_{v_{i}}(\xi_i)$ ($i=1,\ldots,9$) (see text) which give rise to
nine local mode vibrational quantum numbers $v_i$. In order to provide
the conventional normal mode quantum numbers, our local mode
quantum numbers are mapped to their normal mode counterparts by
comparing the corresponding symmetry and polyad of basis functions
with that of the reference (degenerate) harmonic oscillator functions.
The normal mode quantum numbers are $n_1, n_2, L_2, n_3, L_3, M_3,
n_4, L_4, M_4$. The classification of the normal modes under \Td\
symmetry is given in Table~\ref{t:irreps}. The quantum numbers $n_2$
associated with an isotropic 2D harmonic oscillator $|n_2,l_2\rangle $
($E$ symmetry) is complemented with the projection of the vibrational
angular momentum $l_2 = -n_2, -n_2+2 , \ldots, n_2 - 2, n_2$
\citep{04BuJexx.method}. Since the definition of the sign of $l_2$ is
ambiguous (as ambiguous the sign of $k$), we follow the suggestion of
\citet{jt546} and use the absolute value $L_2 = |l_2| = n_2, n_2-2 ,
\ldots, 0$ (1) instead. Similarly, the classification of the 3D normal
modes 3 and 4 is based on the 3D isotropic harmonic oscillator
$|n_i,l_i \rangle$ \citep{01Hoxxxx.CH4} ($i=3,4$), where $n_i$ is the
vibrational quantum number and $l_i$ is the vibrational angular
momentum. Following the classification by \citet{06BoReLo.CH4} we add
the multiplicity index $M_i \le L_i$ which we associate with the
symmetry according with Table~\ref{t:F:M}, where $M_i$ is odd for
$F_i$ and $F_2$ and even for $A_1$, $A_2$, $E$. Imposing the condition
$M_i\le L_i$ automatically constrains $M_i = 0$ as $A_1$, $M_i = 1$
as $F_2$, $M_i = 2$ as $ E$, and $M_i = 3$ as $F_2$. For higher values
of $M_i$ these symmetry designations are repeated as shown in
Table~\ref{t:F:M}.
We also choose
$M_i = 6 + 12 n$ for $A_2$ and $M_i = 12 n $ for $A_1$. It should be
noted however that for $A_1$ and $A_2$ these variations of $M_i$ and
$L_i$ produce over-sampled sets and we choose the combinations with
smallest $M_i$.


Finally, we stress that our stretching vibrational basis set is not
based on the harmonic oscillator functions. We use harmonic
oscillators only as a reference to correlate the local mode quantum
numbers with the conventional normal mode ones.

\label{lastpage}

\end{document}